\documentclass[sigconf,nonacm]{acmart}

\acmBooktitle{}                  

\renewcommand\footnotetextcopyrightpermission[1]{}

\usepackage{array} 
\usepackage{colortbl} 
\usepackage{makecell} 
\usepackage{graphicx}
\usepackage{tcolorbox}
\usepackage{balance}

\AtBeginDocument{%
  }

\begin{document}

\title{Large Language Model Empowered Privacy-Protected Framework for PHI Annotation in Clinical Notes}

\author{Guanchen Wu}
\affiliation{%
  \department{Department of Computer Science}
  \institution{Emory University}
  \city{Atlanta}
  \country{USA}}
\email{guanchen.wu@emory.edu}

\author{Linzhi Zheng}
\affiliation{%
  \department{Department of Computer Science}
  \institution{University of Chicago}
  \city{Chicago}
  \country{USA}}
\email{zhenglz21@mails.tsinghua.edu.cn}

\author{Han Xie}
\affiliation{%
  \department{Department of Computer Science}
  \institution{Emory University}
  \city{Atlanta}
  \country{USA}}
\email{han.xie@emory.edu}

\author{Zhen Xiang}
\affiliation{%
  \department{School of Computing}
  \institution{University of Georgia}
  \city{Atlanta}
  \country{USA}}
\email{zhen.xiang.lance@gmail.com}

\author{Jiaying Lu}
\affiliation{%
  \department{Department of Computer Science}
  \institution{Emory University}
  \city{Atlanta}
  \country{USA}}
\email{jiaying.lu@emory.edu}

\author{Darren Liu}
\affiliation{%
  \department{Nell Hodgson Woodruff School of Nursing}
  \institution{Emory University}
  \city{Atlanta}
  \country{USA}}
\email{darren.liu@emory.edu}

\author{Delgersuren Bold}
\affiliation{%
  \department{Nell Hodgson Woodruff School of Nursing}
  \institution{Emory University}
  \city{Atlanta}
  \country{USA}}
\email{delgersuren.bold@emory.edu}

\author{Bo Li}
\affiliation{%
  \department{Department of Computer Science}
  \institution{University of Chicago}
  \city{Chicago}
  \country{USA}}
\email{lxbosky@gmail.com}

\author{Xiao Hu}
\affiliation{%
  \department{Nell Hodgson Woodruff School of Nursing}
  \institution{Emory University}
  \city{Atlanta}
  \country{USA}}
\email{xiao.hu@emory.edu}

\author{Carl Yang\textsuperscript{\textasteriskcentered}}
\affiliation{%
  \department{Department of Computer Science}
  \institution{Emory University}
  \city{Atlanta}
  \country{USA}}
\email{j.carlyang@emory.edu}

\renewcommand{\shortauthors}{Wu et al.}

\begin{abstract}
The de-identification of private information in medical data is a crucial process to mitigate the risk of confidentiality breaches, particularly when patient personal details are not adequately removed before the release of medical records. Although rule-based and learning-based methods have been proposed, they often struggle with limited generalizability and require substantial amounts of annotated data for effective performance. Recent advancements in large language models (LLMs) have shown significant promise in addressing these issues due to their superior language comprehension capabilities. However, LLMs present challenges, including potential privacy risks when using commercial LLM APIs and high computational costs for deploying open-source LLMs locally. In this work, we introduce LPPA, an \underline{\textbf{L}}LM-empowered \underline{\textbf{P}}rivacy-protected \underline{\textbf{P}}HI \underline{\textbf{A}}nnotation framework for clinical notes, targeting the English language. By fine-tuning LLMs locally with synthetic notes, LPPA ensures strong privacy protection and high PHI annotation accuracy. Extensive experiments demonstrate LPPA's effectiveness in accurately de-identifying private information, offering a scalable and efficient solution for enhancing patient privacy protection.
\end{abstract}





\maketitle

\renewcommand{\thefootnote}{\fnsymbol{footnote}}  
\footnotetext[1]{Corresponding author: j.carlyang@emory.edu}

\section{Introduction}
Clinical notes~\cite{boag2018s} are unstructured text records that document patient encounters during healthcare, commonly including types such as discharge notes, nursing notes, ECG reports, and radiology reports. Discharge notes, in particular, provide detailed accounts of a patient’s hospital stay, capturing physicians' observations, patient interactions, social and behavior determinants of health~\cite{fu2024extracting}, and other clinical nuances. These details go beyond the commonly used structured information (\textit{e.g.} diagnosis, medication, procedures, etc.) stored in Electronic Health Records (EHR), offering insights that are critical for understanding patient conditions and optimizing treatment decisions. The increased sharing of clinical notes holds significant potential to advance healthcare by facilitating more comprehensive studies of disease patterns and accelerating data-driven research through access to richer and more detailed patient data. 
In this work, we utilize the most comprehensive discharge notes as the major data source, but our proposed methods can be generalized to other types of clinical notes.

\begin{figure}[t!]
\centering
\vspace{10pt}
\includegraphics[width=\columnwidth]{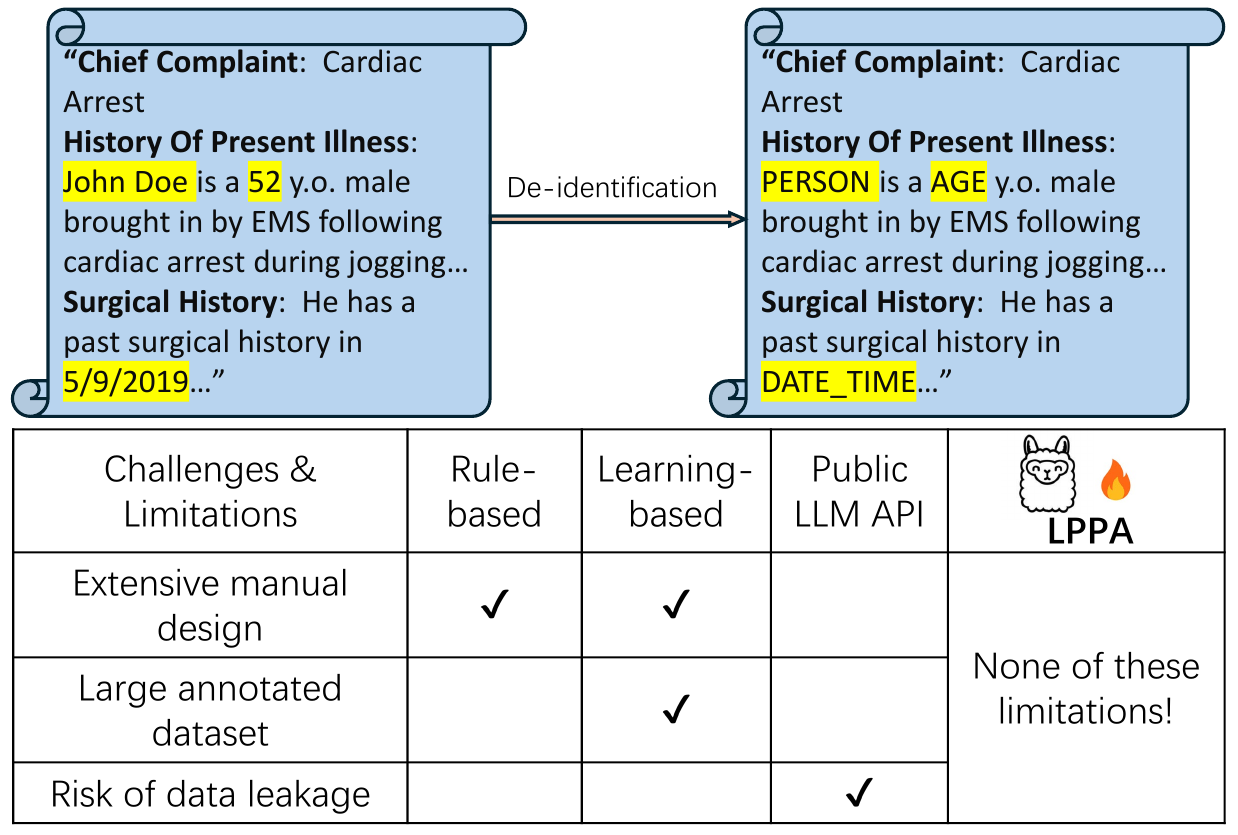}
\caption{Illustration of existing de-identification methods and our proposed de-identification framework.}
\label{fig:1}
\vspace{-10pt}
\end{figure}

However, the sharing of clinical notes across medical and research institutions faces a critical challenge due to patient privacy, as these notes often contain highly sensitive Protected Health Information (PHI)~\cite{moore2019review}. PHI includes identifiable details such as patient names and Social Security Numbers, which can be used to trace specific individuals. For instance, a clinical note might read: ``\textit{John Doe, a 45-year-old male, was diagnosed with hypertension on May 3, 2023}'' where the name, age, and diagnosis date are all PHI. In accordance with the Health Insurance Portability and Accountability Act (HIPAA)~\cite{cohen2018hipaa}, PHI must be thoroughly removed or masked before clinical notes can be shared, posing significant challenges in efficiently sharing the high-utility clinical notes at scale.

Most existing methods for de-identifying private information in text utilize sequence tagging techniques, ranging from rule-based~\cite{gupta2004evaluation} to learning-based approaches~\cite{guo2006identifying,li2016learning}, each with limitations. Rule-based systems, which rely on predefined patterns, struggle with the variability and complexity of clinical text, limiting their scalability. Learning-based methods, including traditional machine learning models like Support Vector Machines (SVMs) and Conditional Random Fields (CRFs), as well as deep learning models such as Recurrent Neural Networks (RNNs) and Long Short-Term Memory networks (LSTMs), improve performance by learning features from data. However, they require large annotated datasets, manual feature selection, and may risk overfitting with limited data. More recently, pre-trained language models (PLMs) like BERT and large language models (LLMs) such as GPT have shown promise in PHI annotation \citep{rezayi2022clinicalradiobert,liu2023deid}, but they face challenges with domain-specific language, high computational costs, and data privacy concerns, especially in healthcare under regulations like HIPAA.

In this work, we propose a framework of \underline{L}LM-empowered \underline{P}rivacy-protected \underline{P}HI \underline{A}nnotation (LPPA), designed to address key limitations of existing de-identification methods, including data scarcity and privacy concerns. Our framework offers two significant contributions: 1) it leverages the flexibility of pre-trained LLMs and few-shot learning to generate synthetic clinical notes, thereby eliminating the need for manual feature engineering and large annotated datasets while capturing the language complexity of real-world clinical notes, and 2) it ensures robust data privacy by fine-tuning a locally hosted LLM for PHI annotation, thus mitigating reliance on external APIs and reliably protecting sensitive information.

We evaluate the proposed framework on both real-world and synthetic clinical notes, benchmarking its performance against a rule-based approach and various LLMs. Our method achieves an F1 score of 0.57 on the real-world clinical note dataset, closely approaching the performance of state-of-the-art LLMs. This result underscores the framework's comparable accuracy while offering additional advantages, such as greater efficiency, scalability, and almost-zero reliance on annotated datasets, making it a practical and privacy-conscious solution for PHI annotation.

\section{Related Work}
Current approaches to de-identification of medical records span from traditional rule-based systems to learning-based methods and more recent LLMs, each facing unique limitations and challenges.

\noindent\textbf{Rule-Based Systems.} Traditional approaches for private information annotation, such as rule-based systems, rely on predefined patterns and domain-specific dictionaries for PHI extraction\citep{gupta2004evaluation,uzuner2007evaluating}. The PhysioNet de-identification software package \citep{neamatullah2008automated}, for instance, uses lexical look-up tables, regular expressions, and simple heuristics to identify PHI in free-text medical records. This system demonstrated high recall and precision but remains limited in scalability and adaptability to unstructured clinical narratives. While effective for structured text, rule-based methods often falter when dealing with the variability and complexity of unstructured clinical narratives, leading to inconsistent results. Additionally, maintaining and scaling these systems, particularly across datasets in different languages, is labor-intensive and inefficient.

\noindent\textbf{Learning-Based Methods.} Machine learning techniques, such as SVMs and CRFs, address some limitations by learning features from annotated datasets \citep{guo2006identifying, li2008conditional}. However, SVMs struggle with high-dimensional feature spaces, and while CRFs are more effective for sequential data, they require extensive manual feature engineering. Additionally, both models often fail to generalize across domains due to variations in data annotation and structure, limiting their broader applicability. Deep learning models, including RNNs \citep{li2016learning, dernoncourt2017identification} and LSTMs \citep{khin2018deep, madan2018redaction}, advance PHI annotation by automatically extracting features, reducing reliance on manual feature selection. However, they also require large annotated datasets to prevent overfitting. Moreover, the computational resources needed for training and inference add complexity, making these models difficult to scale in resource-constrained environments. The 2014 i2b2/UTHealth shared task demonstrated the effectiveness of hybrid approaches combining CRFs with rule-based systems\citep{stubbs2015automated}. This task focused on de-identifying longitudinal clinical narratives, expanding PHI categories beyond HIPAA standards to include professions and full dates for enhanced security. While highlighting progress, the task also underscored challenges in balancing effective de-identification with preserving clinical value. The i2b2 dataset used in the task is not publicly available. \footnote{For more information on the i2b2 dataset, visit \url{https://portal.dbmi.hms.harvard.edu/projects/n2c2-nlp/}.}

\noindent\textbf{Language Models. } More recently, PLMs such as BERT and LLMs such as GPT have shown significant improvements in handling complex language patterns in PHI annotation tasks \citep{liu2023deid,yashwanth2024zero}. These models offer superior contextual understanding, generalization across diverse datasets, and the ability to perform few-shot learning, significantly reducing the need for extensive manual feature engineering and large annotated datasets\citep{wu2024ontology}. Their adaptability and high accuracy make them particularly effective for de-identification tasks compared to traditional rule-based or learning-based methods. However, two key challenges remain: smaller-parameter LLMs often fail to achieve competitive performance, while larger-parameter models, although more effective, demand substantial computational resources for local deployment. Alternatively, using public APIs to mitigate these computational constraints introduces the risk of sensitive data leakage, posing privacy concerns in healthcare settings where security and regulatory compliance are paramount.

\section{Method}
\subsection{Problem Formulation}
Our objective is to accurately detect and annotate PHI entities within clinical notes and replace them with corresponding entity types while minimizing the need for large amounts of manually annotated training data. The de-identification problem is formally defined as follows:

\textbf{Input:}
\begin{itemize}
    \item A clinical note, which contains PHI entities, including names, addresses, dates of birth, and phone numbers.
\end{itemize}

\textbf{Output:}
\begin{itemize}
    \item A fully de-identified clinical note, where each instance of a PHI entity has been identified and replaced with its corresponding entity type label (e.g., "NAME," "ADDRESS," "AGE"). 
\end{itemize}

\subsection{Framework Overview}

In this work, we proposed \underline{\textbf{L}}LM-empowered \underline{\textbf{P}}rivacy-protected \underline{\textbf{P}}HI \underline{\textbf{A}}nnotation framework (LPPA), which addresses key challenges in existing methods for PHI annotation. As illustrated in Figure \ref{fig:1}, rule-based approaches require extensive manual pattern design, learning-based methods depend on large annotated datasets, and utilizing public LLMs through APIs poses significant risks of data exposure. Our framework overcomes these limitations by ensuring high accuracy and low cost while protecting sensitive information and preventing data leaks.

\begin{figure*}[t!]
\centering
\includegraphics[scale=0.47]{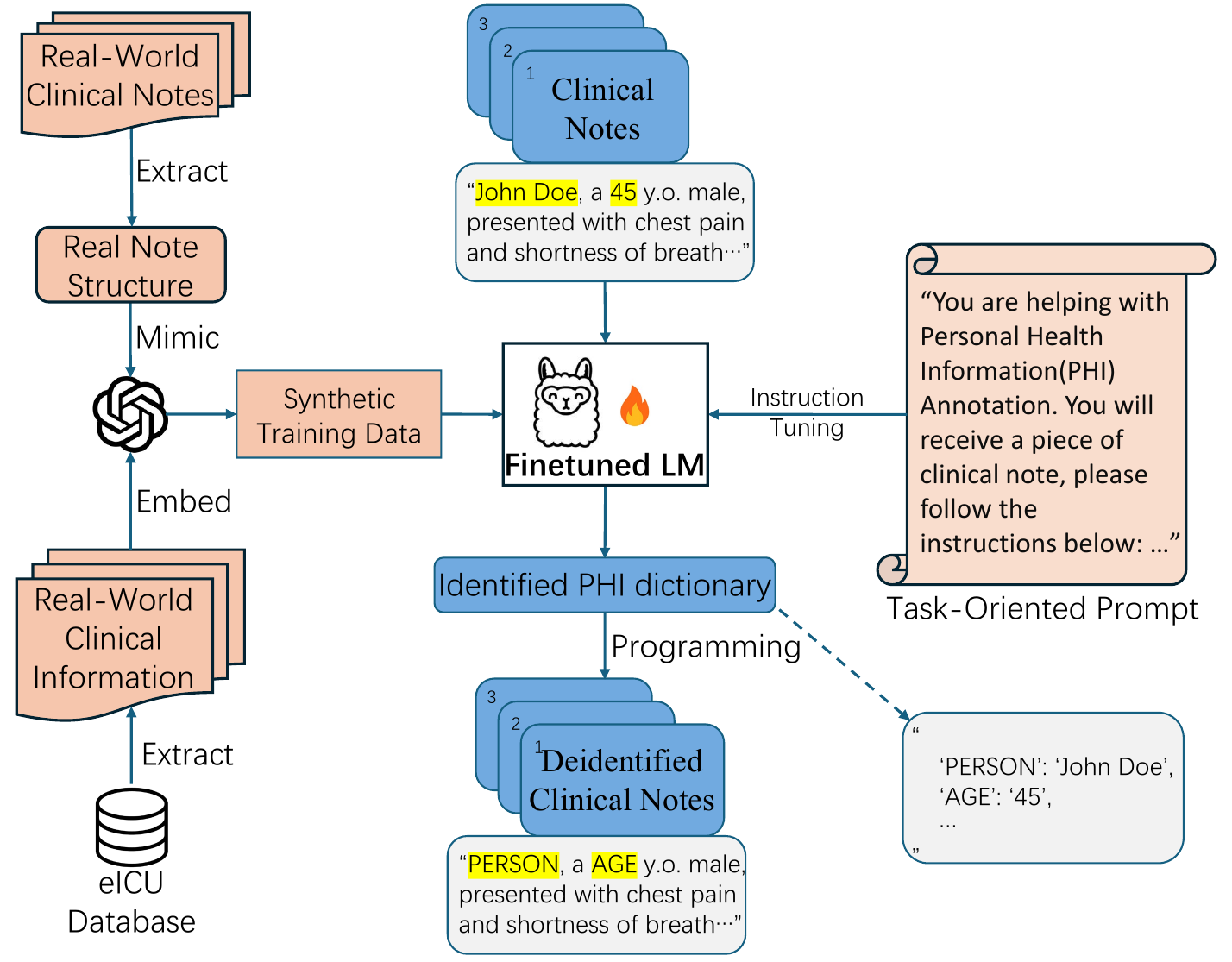}
\caption{LPPA Framework Overview. This framework leverages LLMs to generate synthetic training data using two distinct approaches, which are then combined with a task-oriented prompt to fine-tune a base model. The fine-tuned model processes real clinical notes to output a PHI dictionary identifying sensitive information. A subsequent programming technique is applied to the identified PHI to generate de-identified clinical notes, ensuring privacy while preserving the note's structure.}
\label{fig:2}
\end{figure*}

As shown in Figure \ref{fig:2}, the LPPA framework incorporates two synthetic data generation pipelines—\textbf{Anonymized Example-Guided Note Generation} and \textbf{Synthetic PHI Insertion into Anonymized Notes}—designed to address data scarcity concerns by utilizing different data sources. We also developed a synthetic data mixture pipeline to enhance data diversity and integrated an instruction-tuning process to fine-tune a local language model. This approach eliminates the need for manual feature engineering, reduces dependency on large annotated datasets, and mitigates privacy risks by operating locally without relying on external APIs.

\subsection{LLM-based Synthetic Data Generation} \label{sec:syntheticdataapp1}
Given the limited availability of real-world clinical notes due to strict privacy regulations, synthetic data generation using LLMs plays a crucial role in augmenting the training data for the instruction tuning process in our proposed framework. Recent studies, such as \cite{hiebel2023can}, have demonstrated the effectiveness of synthetic text for clinical NER tasks, highlighting its potential to address data scarcity challenges in medical domains. The primary reason for employing LLMs to generate synthetic data stems from the restricted access to real clinical notes, which are available in limited numbers. Although we possess a small dataset of fully annotated real notes, it is insufficient to train a robust model. Furthermore, publicly available datasets are anonymized, lacking the richness of real-world PHI. By utilizing LLMs, we aim to generate realistic, high-quality synthetic clinical notes that closely resemble real-world notes, thereby significantly expanding our dataset while ensuring compliance with privacy regulations and safeguarding sensitive patient information.

Our synthetic data generation methodology is built upon two key strategies, both focused on producing clinical notes that are structurally and contextually aligned with real-world examples. By harnessing the language generation capabilities of LLMs, we generate synthetic clinical notes that incorporate diverse PHI entities, effectively mimicking the variability and complexity present in real-world clinical data.

\noindent\textbf{Anonymized Example-Guided Note Generation.}
The first approach utilizes a few-shot prompting technique, where the LLM is provided with a small set of fully anonymized, representative real-world clinical notes. Due to limited access to such notes, a carefully curated selection of examples is chosen to help the LLM learn the structure of authentic clinical documentation. These examples, which include essential sections such as patient demographics, medical history, diagnoses, and treatments, serve as a foundation for generating new clinical notes. By leveraging these anonymized examples, the LLM produces synthetic notes that closely replicate the organization and content of real clinical documentation. Simulated PHI entities are seamlessly embedded into the generated notes, ensuring the synthetic data maintains the authenticity and coherence of real notes while strictly adhering to privacy standards by avoiding any reliance on actual patient information.

\noindent\textbf{Synthetic PHI Insertion into Anonymized Notes.} In this approach, we leverage a public dataset as a foundation to generate comprehensive synthetic clinical notes that replicate the richness and structure of real-world medical documentation. Real-world clinical information, such as patient demographics, gender, allergies, diagnoses, medications, and lab results, is utilized as a reference for creating realistic medical content. To ensure privacy, personal identifiers—including patient names, phone numbers, and addresses—are simulated using a LLM. These simulated identifiers are generated based on attributes such as gender and are designed to produce diverse and plausible details. Once generated, these identifiers are seamlessly integrated with the extracted clinical information to produce synthetic clinical notes that closely resemble authentic medical records. In addition to generating these notes, the LLM is tasked with extracting PHI entities to validate the accuracy and completeness of the synthetic data. This method enables the creation of high-quality synthetic notes while adhering to privacy regulations and ensuring no actual patient information is exposed.

\noindent\textbf{Synthetic Data Mixture.} 
To ensure that variations in the quality of synthetic training data generated by the two approaches do not affect the overall performance of our framework, we adopted a strategy of randomly mixing the two datasets. This approach mitigates potential biases arising from differences in data quality between the two methods. By integrating the datasets in this manner, we ensured consistent and unbiased model performance, thereby enhancing the robustness of our framework. Additionally, we evaluated the quality of the synthetic notes in Section \ref{sec:syntheticeval}, providing further insights into their impact on the framework's effectiveness and ensuring a comprehensive assessment of our methodology.

\begin{tcolorbox}[colback=blue!5!white,colframe=blue!50!black,title=\small{Task-Oriented Prompt}]
\small 
\textbf{SystemMessage:} ``You are an experienced doctor who helps with PHI annotation.'' \\
\textbf{HumanMessage:} ``You are helping with Personal Health Information(PHI) Annotation. You will receive a piece of clinical note, please follow the instructions below:\\
1. Identify and extract the following entity types:
[\texttt{"PERSON"}, \texttt{"LOCATION"}, \texttt{"ORGANIZATION"}, \texttt{"AGE"}, \texttt{"PHONE\_NUMBER"}, \texttt{"EMAIL"}, \texttt{"DATE\_TIME"}, \texttt{"ZIP"}, \texttt{"PROFESSION"}, \texttt{"USERNAME"}, \texttt{"ID"}, \texttt{"URL"}]\\
2. Ensure that each identified entity is categorized under the correct entity type from the list above.\\
3. Extract all possible instances of the specified entity types from the clinical note. Even if there is some uncertainty, it's important to include any entity that could potentially belong to one of the listed categories.\\
4. Make sure that the entities identified and extracted are as accurate as possible, but focus on ensuring no relevant entities are missing.\\
5. Your output must be a JSON dictionary where the keys are the specified entity types, and the values are lists of the corresponding identified entities. No explanation needed.\\
Here is the clinical note:'' + Clinical Note
\end{tcolorbox}
\vspace{-1mm}

\subsection{Instruction Tuning}
The primary objective of the PHI annotation task is to accurately and comprehensively extract all possible private information from the given clinical notes. This task requires meticulous identification of every PHI entity present in the text, as any overlooked or missed entity could lead to serious privacy breaches and regulatory compliance failures, particularly in fields governed by strict privacy laws such as healthcare. Therefore, ensuring that our model excels in terms of recall—the metric that measures the model's ability to retrieve all relevant entities from the dataset—is paramount. High recall ensures that the model effectively identifies and extracts every instance of sensitive PHI, significantly reducing the risk of leaving any information exposed or unprotected. A failure to achieve high recall could result in undetected PHI, posing potential legal and ethical risks, including violations of privacy regulations such as the HIPPA. Consequently, recall becomes one of the most critical evaluation metrics for this task, as it directly impacts the model's ability to provide a secure and compliant de-identification solution.

In light of the importance of maximizing recall, we carefully designed the prompts used during the model's fine-tuning process. These prompts are carefully designed to guide the model in prioritizing the identification of all potential PHI entities, minimizing the likelihood of omissions. By instructing the model to prioritize the comprehensive extraction of all relevant PHI entities, we aim to ensure that it can perform well in environments where missing even a single PHI entity could result in significant privacy and compliance issues. Our fine-tuning approach, coupled with prompt engineering, is thus critical in shaping the model to meet the high standards required for this sensitive task, making it not only effective but also reliable in real-world usage.

\section{Experiments}
\subsection{Real-World Datasets}
Due to the sensitive nature of patient information, real-world clinical notes are subject to strict privacy regulations, which significantly limit the number of such documents we can access. In this study, we were able to obtain a collection of 100 fully annotated real-world clinical notes (with all PHI identified and randomly scrambled), generously provided by a large hospital in the US. The usage of these data has been approved under IRB number xxx. Each clinical note averages approximately 1,000 tokens, showcasing a high level of detail and comprehensiveness in documenting patient care. To ensure the integrity and accuracy of the data, several experienced medical experts carefully reviewed and annotated the clinical notes. These experts worked meticulously to identify and scramble ground-truth PHI entities, creating a high-quality dataset for use in our project. Given the inherent limitations on access to real-world clinical notes due to privacy concerns, the size of our dataset remains constrained. As a result, we have opted to use these real-world clinical notes exclusively for evaluation purposes, acknowledging that their limited availability restricts their broader application in the model training phase.

We also utilize datasets from the eICU Collaborative Research Database\citep{pollard2018eicu}, a large public database containing de-identified health records from over 200,000 ICU admissions across multiple hospitals in the United States. The eICU dataset includes a wide range of clinical information, such as patient demographics, diagnoses, treatments, lab results, and medications, making it a valuable resource for research in critical care and healthcare analytics.

\subsection{Experiment Setup}
\noindent\textbf{Evaluation Datasets.} 
Given our limited access to 100 real-world annotated clinical notes, we evaluated our fine-tuned model using this small dataset. To provide a more comprehensive assessment of our proposed framework, we generated an additional evaluation dataset using synthetic data, following the two approaches outlined in Section \ref{sec:syntheticdataapp1}. Specifically, we generated 500 synthetic clinical notes using the real-world data based approach and another 500 synthetic notes using the anonymized data based approach, each accompanied by the corresponding ground truth PHI entities. This synthetic dataset, along with the real notes, enables a thorough evaluation of the model's performance on both real and synthetic data, ensuring a more robust assessment.

\noindent\textbf{Baseline Methods.} 
In this study, we utilized a range of baselines for PHI annotation, including a rule-based approach implemented through the PhysioNet de-identification software package \cite{neamatullah2008automated}. This package employs a combination of regular expressions and lookup dictionaries to identify and replace PHI in unstructured clinical text. We opted not to include traditional learning-based methods in our comparison, as these approaches generally rely on extensive annotated training datasets, which were not available for this work. Due to the limited availability of annotated clinical notes, we deemed learning-based methods unsuitable for the scope of this evaluation. Instead, for the rule-based method, we developed a series of tailored regular expressions to extract specific PHI entities, leveraging the structured patterns often found in clinical documentation.

We also adopted a list of LLMs for comparison. Specifically, we utilized two families of LLMs, i.e., Llama-3-8B-Instruct(which is also the base model we used for the instruction tuning process in this work) and Llama-3-70B-Instruct from Meta's Llama model, along with gpt-3.5-turbo-0301 and gpt-4-0613 from OpenAI's GPT model. Note that since we evaluate these baseline models using real-world clinical notes, experiments need to be conducted locally so that we can avoid private data from leaking. Meta's Llama models we used in this work are open-source LLMs, and hence can be downloaded locally and thus no privacy issues. However, directly using OpenAI's API would potentially leak patients' private data. Hence, we used Microsoft's Azure platform to access the two GPT models we used. Using the Azure platform does not pose data privacy concerns as Microsoft ensures prompt data is not stored after processing, and the platform complies with strict security and regulatory standards such as encryption.

\noindent\textbf{Evaluation Metrics.} 
In this study, we evaluate the performance of our PHI annotation model using three key metrics: precision, recall, and F1-score. These metrics are widely recognized in the field of information extraction and provide a comprehensive assessment of the model's ability to accurately and thoroughly identify PHI entities in clinical notes. Precision measures the accuracy of the model’s predictions, ensuring that the identified PHI entities are correct and relevant. A high precision value indicates fewer false positives, which is crucial for minimizing incorrect annotations that could lead to misidentification or unnecessary redaction of non-PHI data. In PHI annotation tasks, precision is essential to maintain the integrity of clinical documentation, as over-annotating non-sensitive information could distort the meaning of the text.

Recall, in contrast, evaluates the model’s ability to identify all relevant PHI entities, ensuring that no sensitive information is missed. High recall is crucial to prevent the omission of PHI, which could lead to privacy breaches and non-compliance with regulations such as HIPAA. Striking a balance between precision and recall is essential, as focusing too much on precision can result in missed PHI entities, while overemphasizing recall may introduce false positives, incorrectly labeling non-PHI data as sensitive. Hence we also use the F1-score, which balances precision and recall and provides a useful measure of overall performance. Together, these metrics offer a comprehensive evaluation framework, helping to assess the trade-offs between accurately identifying PHI entities and minimizing errors, which is critical for effective PHI annotation in real-world clinical settings.

\subsection{Experimental Implementation}

During the instruction tuning process, we generated 4,000 synthetic clinical notes using the gpt-4-0125-preview model, leveraging two distinct approaches illustrated in Section \ref{sec:syntheticdataapp1}. Specifically, the real-world data-based approach produced 3,000 synthetic notes, while the anonymized data-based approach generated an additional 1,000 notes. Further details on the data generation process are available in Appendix~\ref{appendix:trainingdata}, and specific synthetic data generation prompt are provided in Appendix~\ref{appendix:approach2}.

\begin{table*}
\centering
\renewcommand{\arraystretch}{1.5}
\resizebox{\textwidth}{!}{
\begin{tabular}{|c|ccc|ccc|ccc|ccc|}
\hline
\textbf{Models}     & \multicolumn{3}{c|}{\textbf{Overall}}                                                                                                                                                           & \multicolumn{3}{c|}{\textbf{PERSON}}                                                                                                                                                & \multicolumn{3}{c|}{\textbf{AGE}}                                                                                                                                                     & \multicolumn{3}{c|}{\textbf{DATE/TIME}}                                                                                                                                    \\ \hline
                    & \multicolumn{1}{c|}{\textbf{Pr}}                                         & \multicolumn{1}{c|}{\textbf{Re}}                                & \textbf{F1}                                        & \multicolumn{1}{c|}{\textbf{Pr}}                                        & \multicolumn{1}{c|}{\textbf{Re}}                      & \textbf{F1}                                       & \multicolumn{1}{c|}{\textbf{Pr}}                                        & \multicolumn{1}{c|}{\textbf{Re}}                       & \textbf{F1}                                        & \multicolumn{1}{c|}{\textbf{Pr}}                               & \multicolumn{1}{c|}{\textbf{Re}}                               & \textbf{F1}                              \\ \hline
PhysioNet           & \multicolumn{1}{c|}{0.39$^*$}                             & \multicolumn{1}{c|}{0.24$^*$}                    & 0.28$^*$                           & \multicolumn{1}{c|}{0.01$^*$}                            & \multicolumn{1}{c|}{0.20$^*$}          & 0.02$^*$                           & \multicolumn{1}{c|}{/}                                                  & \multicolumn{1}{c|}{/}                                 & /                                                  & \multicolumn{1}{c|}{\cellcolor{red!30}0.91$^*$} & \multicolumn{1}{c|}{\cellcolor{red!30}0.76$^*$} & \cellcolor{red!30}0.83$^*$ \\ \hline
Llama3-8B-Instruct  & \multicolumn{1}{c|}{0.46$\pm$0.02}                                       & \multicolumn{1}{c|}{0.59$\pm$0.03}                              & 0.50$\pm$0.02                                      & \multicolumn{1}{c|}{0.53$\pm$0.01}                                      & \multicolumn{1}{c|}{\cellcolor{blue!30}0.55$\pm$0.02}                    & 0.53$\pm$0.01                                     & \multicolumn{1}{c|}{0.38$\pm$0.05}                                      & \multicolumn{1}{c|}{0.43$\pm$0.02}                     & 0.41$\pm$0.04                                      & \multicolumn{1}{c|}{0.79$\pm$0.03}                             & \multicolumn{1}{c|}{0.39$\pm$0.02}                             & 0.52$\pm$0.01                            \\ \hline
Llama3-70B-Instruct & \multicolumn{1}{c|}{0.60$\pm$0.01$^*$}                    & \multicolumn{1}{c|}{\cellcolor{blue!30}0.68$\pm$0.01$^*$}           & \cellcolor{red!30}0.62$\pm$0.01$^*$ & \multicolumn{1}{c|}{0.59$\pm$0.01$^*$}                   & \multicolumn{1}{c|}{0.53$\pm$0.01}                    & 0.56$\pm$0.01$^*$                  & \multicolumn{1}{c|}{0.48$\pm$0.02$^*$}                   & \multicolumn{1}{c|}{0.42$\pm$0.01}                     & 0.45$\pm$0.01                                      & \multicolumn{1}{c|}{\cellcolor{blue!30}0.83$\pm$0.01}                             & \multicolumn{1}{c|}{0.50$\pm$0.01$^*$}          & 0.63$\pm$0.01$^*$         \\ \hline
GPT-3.5-turbo       & \multicolumn{1}{c|}{0.43$\pm$0.02}                                       & \multicolumn{1}{c|}{0.60$\pm$0.01}                              & 0.48$\pm$0.02                                      & \multicolumn{1}{c|}{0.60$\pm$0.04}                                      & \multicolumn{1}{c|}{0.50$\pm$0.01}                    & 0.54$\pm$0.02                                     & \multicolumn{1}{c|}{0.39$\pm$0.02}                                      & \multicolumn{1}{c|}{0.36$\pm$0.01\textsuperscript{**}} & 0.37$\pm$0.01$^*$                  & \multicolumn{1}{c|}{0.74$\pm$0.03}                             & \multicolumn{1}{c|}{0.44$\pm$0.02\textsuperscript{**}}         & 0.55$\pm$0.02                            \\ \hline
GPT-4               & \multicolumn{1}{c|}{0.53$\pm$0.01$^*$}                    & \multicolumn{1}{c|}{\cellcolor{red!30} 0.69$^*$} & \cellcolor{blue!30}0.58$\pm$0.01$^*$                   & \multicolumn{1}{c|}{\cellcolor{blue!30}0.60$\pm$0.02$^*$}                  & \multicolumn{1}{c|}{\cellcolor{red!30}0.57$\pm$0.01}  & \cellcolor{red!30}0.58$\pm$0.01$^*$ & \multicolumn{1}{c|}{0.45$\pm$0.03}                                      & \multicolumn{1}{c|}{0.37$\pm$0.03}                     & 0.41$\pm$0.03                                      & \multicolumn{1}{c|}{0.80$\pm$0.01}                             & \multicolumn{1}{c|}{\cellcolor{blue!30}0.57$\pm$0.02$^*$}         & \cellcolor{blue!30}0.67$\pm$.02$^*$         \\ \hline
1K AEG Model        & \multicolumn{1}{c|}{0.47$\pm$0.01}                                       & \multicolumn{1}{c|}{0.57$\pm$0.01}                              & 0.50$\pm$0.01                                      & \multicolumn{1}{c|}{0.48$\pm$0.01$^*$}                   & \multicolumn{1}{c|}{0.52$\pm$0.02}                    & 0.50$\pm$0.01$^*$                 & \multicolumn{1}{c|}{0.49$\pm$0.04$^*$}                   & \multicolumn{1}{c|}{\cellcolor{red!30}0.45$\pm$.02}    & 0.47$\pm$0.03                                      & \multicolumn{1}{c|}{0.66$\pm$0.03$^*$}          & \multicolumn{1}{c|}{0.43$\pm$0.01$^*$}          & 0.52$\pm$0.02                            \\ \hline
2K AEG Model        & \multicolumn{1}{c|}{0.54$\pm$0.01$^*$}                    & \multicolumn{1}{c|}{0.61$\pm$0.02}                              & 0.55$\pm$0.01$^*$                   & \multicolumn{1}{c|}{0.54$\pm$0.02}                                      & \multicolumn{1}{c|}{0.53}                             & 0.54$\pm$0.02                                     & \multicolumn{1}{c|}{0.50$\pm$0.02$^*$}                   & \multicolumn{1}{c|}{0.44$\pm$0.02}                     & 0.46$\pm$0.2                                       & \multicolumn{1}{c|}{0.71$\pm$0.05}                             & \multicolumn{1}{c|}{0.44$\pm$0.01$^*$}          & 0.54$\pm$0.02                            \\ \hline
3K AEG Model        & \multicolumn{1}{c|}{0.55$\pm$0.01$^*$}                    & \multicolumn{1}{c|}{0.61$\pm$0.01}                              & 0.56$\pm$0.01$^*$                   & \multicolumn{1}{c|}{0.56$\pm$0.01$^*$}                   & \multicolumn{1}{c|}{0.54$\pm$0.01}                    & 0.55$\pm$0.01                                     & \multicolumn{1}{c|}{0.51$\pm$0.01$^*$} & \multicolumn{1}{c|}{0.41$\pm$0.01}                     & 0.45$\pm$0.01                                      & \multicolumn{1}{c|}{0.73$\pm$0.05}                             & \multicolumn{1}{c|}{0.42$\pm$0.02}                             & 0.53$\pm$0.03                            \\ \hline
1K SPI Model        & \multicolumn{1}{c|}{0.51$\pm$0.01$^*$}                    & \multicolumn{1}{c|}{0.54$^*$}                    & 0.50$\pm$0.01                                      & \multicolumn{1}{c|}{0.58$\pm$0.01$^*$}                   & \multicolumn{1}{c|}{0.54$\pm$0.01}                    & 0.56$\pm$0.01$^*$                  & \multicolumn{1}{c|}{0.48$\pm$0.01$^*$}                   & \multicolumn{1}{c|}{0.42$\pm$0.02}                     & 0.45$\pm$0.01                                      & \multicolumn{1}{c|}{0.72$\pm$0.01$^*$}          & \multicolumn{1}{c|}{0.40$\pm$0.02}                             & 0.51$\pm$0.02                            \\ \hline
2K SPI Model        & \multicolumn{1}{c|}{0.52$\pm$0.02$^*$}                    & \multicolumn{1}{c|}{0.53$\pm$0.02$^*$}           & 0.50$\pm$0.02                                      & \multicolumn{1}{c|}{0.55$\pm$0.02}                                      & \multicolumn{1}{c|}{0.50$\pm$0.01$^*$} & 0.52$\pm$0.01                                     & \multicolumn{1}{c|}{\cellcolor{red!30}0.52$\pm$0.02$^*$}                   & \multicolumn{1}{c|}{0.41$\pm$0.01}                     & 0.46$\pm$0.01                                      & \multicolumn{1}{c|}{0.79$\pm$0.03}                             & \multicolumn{1}{c|}{0.34$\pm$0.01$^*$}          & 0.48$\pm$0.01$^*$         \\ \hline
3K Hybrid Model     & \multicolumn{1}{c|}{0.59$\pm$0.01$^*$}                    & \multicolumn{1}{c|}{0.53$\pm$0.02$^*$}           & 0.53$\pm$0.01$^*$                   & \multicolumn{1}{c|}{\cellcolor{red!30}0.61$\pm$0.03$^*$} & \multicolumn{1}{c|}{0.54$\pm$0.01}                    & \cellcolor{blue!30}0.57$\pm$0.02                                     & \multicolumn{1}{c|}{0.51$\pm$0.02$^*$}                   & \multicolumn{1}{c|}{\cellcolor{blue!30}0.44$\pm$0.01}                     & \cellcolor{blue!30}0.47$\pm$0.02$^*$                   & \multicolumn{1}{c|}{0.76$\pm$.02}                              & \multicolumn{1}{c|}{0.36$\pm$0.01}                             & 0.49$\pm$0.01$^*$         \\ \hline
4K Hybrid Model     & \multicolumn{1}{c|}{\cellcolor{red!30} 0.65$\pm$0.01$^*$} & \multicolumn{1}{c|}{0.54$\pm$0.02}                              & 0.57$\pm$0.01$^*$                   & \multicolumn{1}{c|}{0.59$\pm$0.02$^*$}                   & \multicolumn{1}{c|}{0.53$\pm$0.02}                    & 0.56$\pm$0.02                                     & \multicolumn{1}{c|}{0.49$\pm$0.01$^*$}                   & \multicolumn{1}{c|}{0.42$\pm$0.02}                     & 0.45$\pm$0.01                                      & \multicolumn{1}{c|}{0.82$\pm$0.03}                             & \multicolumn{1}{c|}{0.40$\pm$0.03}                             & 0.54$\pm$0.03                            \\ \hline
5K Hybrid Model     & \multicolumn{1}{c|}{\cellcolor{blue!30}0.64$\pm$0.01$^*$}                    & \multicolumn{1}{c|}{0.55$\pm$0.02}                              & 0.57$\pm$0.01$^*$                   & \multicolumn{1}{c|}{0.59$\pm$0.03$^*$}                   & \multicolumn{1}{c|}{0.53$\pm$0.02}                    & 0.56$\pm$0.01$^*$                  & \multicolumn{1}{c|}{\cellcolor{blue!30}0.51$\pm$0.01$^*$} & \multicolumn{1}{c|}{0.44$\pm$0.02}                     & \cellcolor{red!30}0.48$\pm$0.02$^*$ & \multicolumn{1}{c|}{0.82$\pm$0.02}                             & \multicolumn{1}{c|}{0.35$\pm$0.02}                             & 0.50$\pm$0.01                            \\ \hline
\end{tabular}
}

\caption{This table presents the average evaluation results for different baseline models and our fine-tuned models on 100 real-world clinical notes. The ``1K AEG Model'' was fine-tuned using 1,000 synthetic notes generated through the \underline{\textbf{A}}nonymized \underline{\textbf{E}}xample-\underline{\textbf{G}}uided Note Generation approach. The ``SPI Model'' was fine-tuned with synthetic data generated via the \underline{\textbf{S}}ynthetic \underline{\textbf{P}}HI \underline{\textbf{I}}nsertion approach. ``Hybrid Model'' integrates both tuning methods. The reported scores represent averages across all 100 clinical notes. \textcolor{red!60}{Red} cells indicate the best results, and \textcolor{blue!60}{Blue} cells indicate the second-best results in each metric. The `/' sign indicates that the model cannot identify this PHI category. We conducted paired t-tests between each model and the backbone model, Llama3-8B-Instruct. We use * to indicate statistically significant results (p<0.05).}
\label{tab:1}
\vspace{-8mm} 
\end{table*}

\subsection{Results}

The experimental results, presented in Table \ref{tab:1}, are average scores computed over 100 real-world clinical notes, showcasing the performance of various models on the PHI annotation task. PhysioNet's Deid tool, serves as a rule-based method and demonstrates moderate general performance, excelling in structured categories like DATE/TIME but struggling significantly with unstructured entities such as PERSON and AGE. Among the LLMs, Llama-3-70B-Instruct achieves the highest overall performance, surpassing both its smaller counterpart, Llama-3-8B-Instruct, and GPT-3.5-turbo. This highlights the benefits of increased model size in handling complex PHI categories. GPT-4 also exhibits strong performance, particularly in achieving a balance between precision and recall across various entity types.

The fine-tuned models use synthetic data generated through approaches detailed in Section \ref{sec:syntheticdataapp1}, including the Anonymized Example-Guided Note Generation approach (AEG) and the Synthetic PHI Insertion into Anonymized Notes approach (SPI). Models fine-tuned with SPI data exhibit consistent improvements in recall, while those fine-tuned with AEG data achieve higher precision but slightly lower recall compared to their base model. Hybrid models, which integrate data from both approaches, outperform Llama-3-70B-Instruct in precision and achieve robust general performance across PHI categories. However, the trade-off between precision and recall is evident, as higher precision in these models sometimes comes at the cost of missing relevant entities. This underscores the inherent challenge of balancing strictness and coverage in PHI annotation tasks, where achieving robustness requires optimizing for both precision and recall.

The results of the ablation study within Table \ref{tab:1}, specifically the rows between ``1K AEG Model'' and ``5K Hybrid Model'', further explore the performance of fine-tuned models trained on varying data sizes generated using AEG, SPI, and Hybrid approaches. Models fine-tuned with AEG exhibit consistent improvements in both precision and recall as the data size increases, reaching optimal performance at 3K data points. In comparison, SPI-tuned models show slightly lower overall performance, particularly in recall, with diminishing gains and a marginal decline in general performance as data size grows. Hybrid models achieve the highest precision and F1 score at 4K data points, outperforming both AEG and SPI-tuned models. However, at 5K data points, Hybrid-tuned models experience a decline in precision while F1 remains the same, suggesting a trade-off between these metrics. This indicates that while the Hybrid approach enhances precision, it may reduce recall by overlooking a greater number of relevant entities. 

\begin{table}[t!]
\centering
\renewcommand{\arraystretch}{} 
\resizebox{\columnwidth}{!}{
\begin{tabular}{|c|ccc|}
\hline
\textbf{}             & \multicolumn{3}{c|}{\textbf{Metrics}}                                             \\ \hline
\textbf{Models}       & \multicolumn{1}{c|}{\textbf{Ave.Pr}} & \multicolumn{1}{c|}{\textbf{Ave.Re}} & \textbf{Ave.F1} \\ \hline
PhysioNet Deid        & \multicolumn{1}{c|}{0.46$\pm$0.02$^*$}        & \multicolumn{1}{c|}{0.32$\pm$0.01$^*$}        & 0.36$\pm$0.01$^*$        \\ \hline
Llama-3-8B-Instruct   & \multicolumn{1}{c|}{0.57$\pm$0.04}        & \multicolumn{1}{c|}{0.67$\pm$0.03}        & 0.61$\pm$0.03        \\ \hline
Llama-3-70B-Instruct  & \multicolumn{1}{c|}{0.66$\pm$0.02$^*$}        & \multicolumn{1}{c|}{0.77$\pm$0.02$^*$}        & 0.71$\pm$0.02$^*$        \\ \hline
GPT-3.5-turbo         & \multicolumn{1}{c|}{0.60$\pm$0.01}        & \multicolumn{1}{c|}{0.76$\pm$0.02$^*$}        & 0.66$\pm$0.01        \\ \hline
GPT-4                 & \multicolumn{1}{c|}{0.68$\pm$0.02$^*$}        & \multicolumn{1}{c|}{0.77$\pm$0.03$^*$}        & 0.71$\pm$0.02$^*$        \\ \hline
3k AEG Model    & \multicolumn{1}{c|}{0.83$\pm$0.01$^*$}        & \multicolumn{1}{c|}{0.84$\pm$0.02$^*$}        & 0.83$\pm$0.01$^*$        \\ \hline
2k SPI Model    & \multicolumn{1}{c|}{0.79$\pm$0.02$^*$}        & \multicolumn{1}{c|}{0.69$\pm$0.02}        & 0.73$\pm$0.03$^*$       \\ \hline
4k Hybrid Model & \multicolumn{1}{c|}{0.87$\pm$0.01$^*$}        & \multicolumn{1}{c|}{0.86$\pm$0.01$^*$}        & 0.86$\pm$0.01$^*$      \\ \hline
\end{tabular}
}
\caption{This table compares the evaluation results of baseline models and our fine-tuned model on the synthetic evaluation dataset. The Hybrid Model outperforms both the baseline LLMs and other fine-tuned models. We conducted paired t-tests between each model and the backbone model, Llama3-8B-Instruct. We use $^*$ to indicate statistically significant results (p<0.05).}
\label{tab:2}
\vspace{-12mm}
\end{table}

Table \ref{tab:2} presents the evaluation results on a synthetic dataset containing 1,000 synthetic clinical notes. The overall performance trend mirrors that of Table \ref{tab:1}. Specifically, the Hybrid-Tuned Model consistently outperforms models fine-tuned using only one of the two synthetic data generation approaches. Notably, when evaluated on the synthetic dataset, the fine-tuned models demonstrate superior performance compared to the LLMs. However, when evaluated on real-world clinical notes, the performance of the fine-tuned models is comparable to that of the LLMs, highlighting the influence of data sources on model performance. Given that the Hybrid fine-tuned model was trained on a large amount of synthetic data, its stronger performance on the synthetic dataset is expected.

\subsection{Synthetic Data Evaluation} \label{sec:syntheticeval}

The evaluation of synthetic clinical notes generated by the LLM was conducted using four metrics: Self-BLEU, Perplexity, Entropy, and Medical Plausibility. These metrics collectively assess critical aspects of text quality, including diversity, fluency, lexical richness, and alignment with medical ontologies. Self-BLEU measures diversity within the dataset, with lower scores indicating greater variation among clinical notes and reduced redundancy. This metric reflects the model's ability to produce diverse outputs while avoiding repetitive patterns. Perplexity evaluates fluency by quantifying how predictable the word sequences are according to a pre-trained language model, where lower scores correspond to more coherent and natural text. Entropy captures the lexical richness of the notes, with higher scores reflecting a broader vocabulary and greater linguistic variability, which are essential for accurately modeling complex clinical scenarios. Finally, Medical Plausibility examines the inclusion of valid medical entities by extracting entities from the clinical notes using an NLP model and verifying their alignment with the UMLS ontology. A note is deemed plausible if it contains at least one entity that matches an entry in the ontology, and the overall Medical Plausibility score represents the proportion of such notes within the dataset.

\begin{table}[t]
\renewcommand{\arraystretch}{1.3}
\centering

\scalebox{0.82}{

\begin{tabular}{|c|cccc|}
\hline
                                 & \multicolumn{4}{c|}{\textbf{Metrics}}                                                                                                                             \\ \hline
\textbf{Data}             & \multicolumn{1}{c|}{\textbf{BLEU (\(\downarrow\))}} & \multicolumn{1}{c|}{\textbf{Perp. (\(\downarrow\))}} & \multicolumn{1}{c|}{\textbf{Entr. (\(\uparrow\))}} & \textbf{Plau. (\(\uparrow\))} \\ \hline
Real Notes        & \multicolumn{1}{c|}{0.55}                               & \multicolumn{1}{c|}{37.59}                               & \multicolumn{1}{c|}{11.36}                        & 0.93                                       \\ \hline
AEG Notes           & \multicolumn{1}{c|}{0.60 $\pm$ 0.01$^*$}                               & \multicolumn{1}{c|}{24.40 $\pm$ 0.65$^*$}                               & \multicolumn{1}{c|}{8.65 $\pm$ 0.06$^*$}                         & 0.97 $\pm$ 0.02$^*$                                       \\ \hline
SPI Notes          & \multicolumn{1}{c|}{0.35 $\pm$ 0.01$^*$}                               & \multicolumn{1}{c|}{34.63 $\pm$ 2.56$^*$}                               & \multicolumn{1}{c|}{10.28 $\pm$ 0.08$^*$}                        & 0.92 $\pm$ 0.02$^*$                                    \\ \hline
Mixture Notes & \multicolumn{1}{c|}{0.49 $\pm$ 0.01$^*$}                               & \multicolumn{1}{c|}{30.27 $\pm$ 0.65$^*$}                               & \multicolumn{1}{c|}{9.80 $\pm$ 0.06$^*$}                         & 0.92 $\pm$ 0.02$^*$                                     \\ \hline
\end{tabular}
}
\caption{Evaluation results for real-world and synthetic clinical notes. The data sources include real-world clinical notes(Real Notes), notes generated individually by AEG and SPI, and the mixture of synthetic clinical notes (Mixture Notes, containing notes generated by AEG and SPI). The arrows indicate whether lower (\(\downarrow\)) or higher (\(\uparrow\)) values are preferable for each metric. We conducted paired t-tests between each data source and Real Notes. We use $^*$ to indicate statistically significant results (p<0.01).}
\label{tab:3}
\vspace{-10mm}
\end{table}

The results, presented in Table~\ref{tab:3}, indicate that synthetic clinical notes generated using different methods exhibit varying performance across the evaluated metrics, highlighting trade-offs in quality dimensions. Real-world clinical notes achieve a balanced performance with moderate diversity (Self-BLEU), high lexical richness (Entropy), and strong inclusion of valid medical entities (Medical Plausibility), though they exhibit higher Perplexity, reflecting the complexity and unpredictability of authentic clinical language. The Mixture Synthetic Clinical Notes closely approximate the real-world notes, with slightly improved fluency (lower Perplexity) but a reduction in lexical richness (lower Entropy). Among the synthetic methods, notes generated by AEG demonstrates the best fluency and the highest Medical Plausibility score, suggesting its effectiveness in generating coherent text that aligns strongly with medical ontologies; however, it has lower diversity (higher Self-BLEU) and limited lexical variety (lower Entropy), which may indicate repetitive outputs. In contrast, notes generated by SPI exhibits the highest diversity and improved lexical richness compared to other synthetic methods, yet it has slightly higher Perplexity, indicating a trade-off between diversity and fluency. Since AEG and SPI have different quality profiles, future work can focus on refining synthetic data generation by leveraging these evaluations to address their respective weaknesses. By improving diversity, fluency, or richness in targeted ways, the synthetic data can be further optimized, potentially enabling more robust model performance and better alignment with real-world clinical note characteristics.

\subsection{Case Studies}
\begin{figure}[t!]
\centering
\includegraphics[width=\columnwidth]{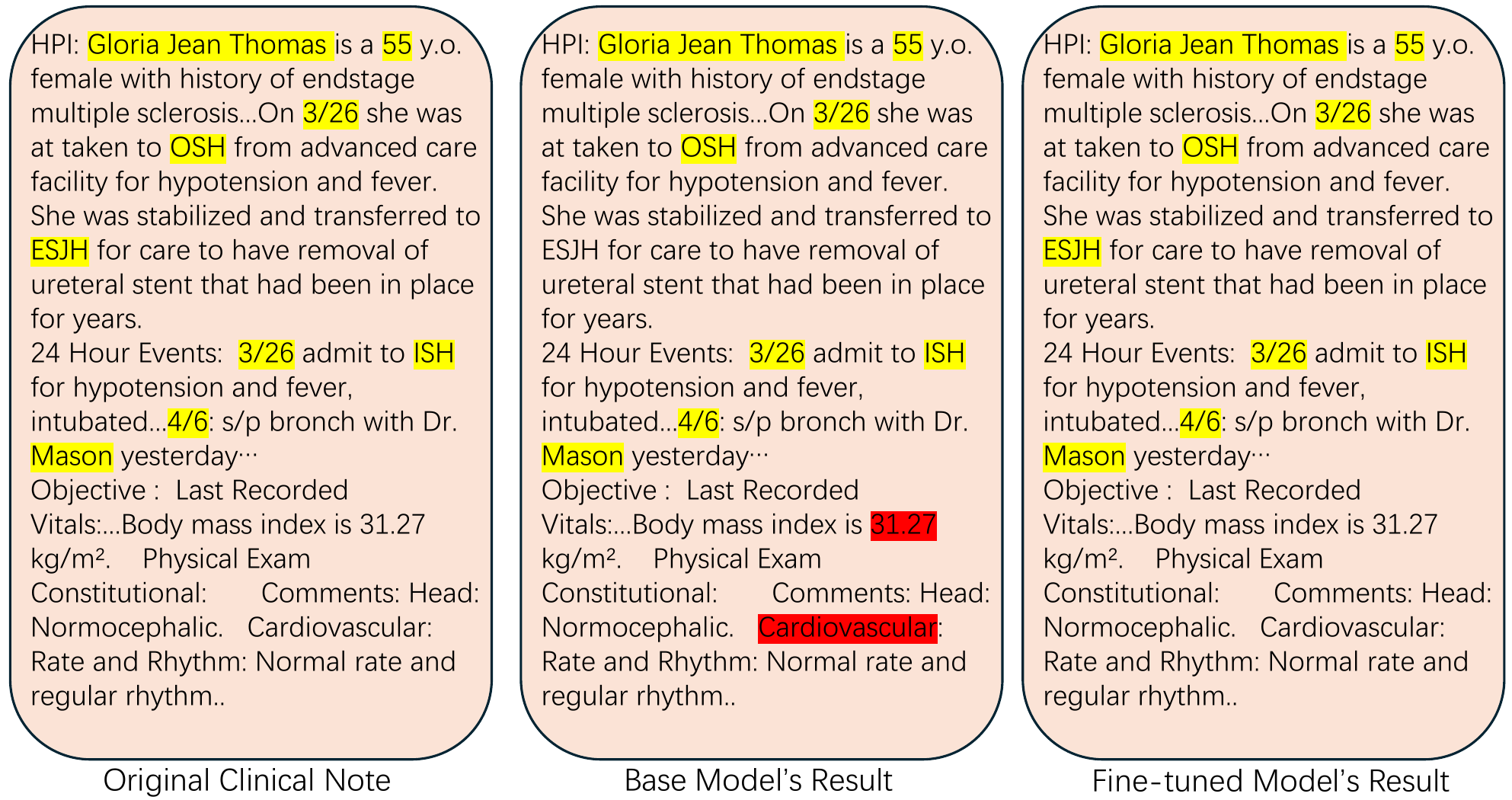}
\caption{Case Study}
\label{fig:3}

\end{figure}

To provide a more detailed analysis, we present a case study, showcasing both its original and de-identified versions. The de-identification outcomes from both the base model and our fine-tuned model are compared to highlight their relative performance. Correctly identified PHI entities are highlighted in yellow. As depicted in Figure \ref{fig:3}, the base model incorrectly identifies several PHI entities (highlighted in red), while our fine-tuned model accurately annotates all correct PHI entities. This case study underscores the enhanced robustness and accuracy of our proposed framework when applied to real-world clinical data, demonstrating its superior ability to handle complex de-identification tasks.

\subsection{Computational and Financial Costs} 
Estimating the computational costs of our proposed framework is crucial, with the instruction tuning process consuming the most significant resources. In this work, the fine-tuning task was performed on a local server equipped with eight H100 GPUs. For the fine-tuning task, we utilized two of these H100 GPUs. Our best-performing fine-tuned model was trained using 4,000 synthetic data samples, with the entire fine-tuning process taking approximately 10 minutes to complete. Further information on the fine-tuning hyperparameters can be found in Appendix~\ref{appendix:hyper}, and more details about the fine-tuning process are provided in the Appendix~\ref{appendix:finetune}.

\begin{figure}[t!]
\centering
\includegraphics[width=\columnwidth,height=8cm,keepaspectratio]{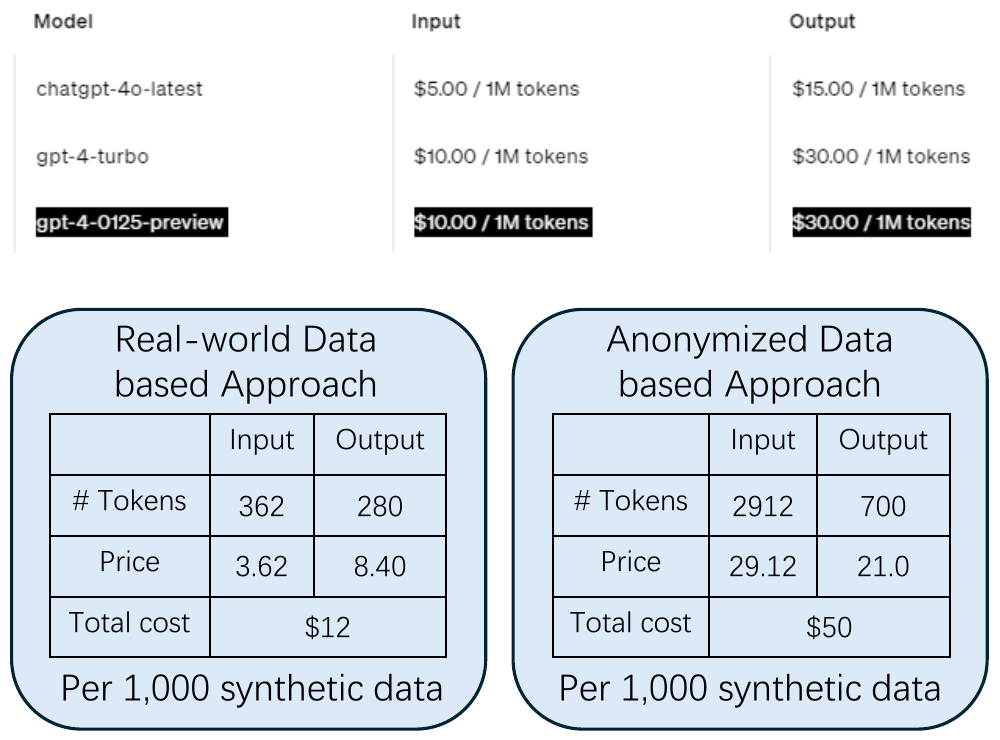}
\caption{Cost Estimation}
\label{fig:4}
\vspace{-5mm}
\end{figure}

Regarding financial costs, the fine-tuning task itself incurred no expenses, as it was executed on a local server. However, synthetic data generation, a key component of our framework, involved using GPT models via OpenAI's API. Specifically, we generated 4,000 synthetic data samples for both fine-tuning and evaluation, with each API call involving an average of 1,600 tokens for input and 500 tokens for output. A detailed breakdown of the cost estimation is provided in Figure \ref{fig:4}.

\section{Conclusions}
In this work, we introduced an innovative LLM-empowered Privacy-Protected PHI Annotation (LPPA) framework designed to address key limitations of existing methods for PHI de-identification in clinical notes. By leveraging synthetic data generation, our approach bypasses the need for large annotated datasets, while instruction-tuning ensures high performance in identifying sensitive private information. Our experimental results demonstrated that the fine-tuned models significantly improve PHI annotation performance, achieving competitive accuracy with minimal reliance on manually annotated data. The framework effectively balances privacy, scalability, and computational efficiency, making it a practical solution for safeguarding patient data in real-world clinical settings.

The broader applications of this framework extend beyond healthcare, with potential usage in any domain requiring privacy-preserving text analysis, such as legal documents or financial records. Future work can explore further enhancements by integrating multimodal data sources or improving the generalizability of the model to handle more diverse types of PHI. Additionally, refining synthetic data based on evaluation metrics can further improve model performance, enhancing its effectiveness and reliability. Optimizing the balance between computational cost and model performance while adhering to privacy regulations would also enable wider adoption of the LPPA framework in diverse regulatory environments.

\appendix

\section{Fine-tuning Hyper Parameters}
\label{appendix:hyper}
In this study, we fine-tuned an LLM taking Llama-3-8B-Instruct as the base model, employing the LoRA (Low-Rank Adaptation) method to reduce computational overhead while maintaining high performance. The model was optimized using the AdamW optimizer with a fixed learning rate of $1 \times 10^{-4}$, and gradient accumulation was set to 2 steps to enable efficient training on GPU resources. LoRA was configured with a rank of 16 and an alpha value of 32, ensuring an optimal balance between model capacity and computational efficiency. A dropout rate of 0.05 was applied to prevent overfitting, and specific projection layers and attention heads were targeted for fine-tuning. Text preprocessing included tokenizing clinical notes and structuring them in a chat format, with user inputs corresponding to the clinical notes and assistant outputs representing the annotated PHI, ensuring a conversational model training setup.

The model was trained on the synthetic dataset with a batch size of 8 for training and 1 for evaluation, while a sequence length of 512 tokens was maintained to capture sufficient context from the clinical notes. Mixed-precision training was employed by using the bfloat16 data type to optimize memory consumption without compromising numerical stability. Evaluations were performed at regular intervals during training, with \texttt{eval\_steps} set to 0.25 of an epoch, and training logs were captured at every step to monitor model performance. To further enhance efficiency, input sequences were grouped by length during training, minimizing padding and improving computational resource utilization. These hyperparameters collectively ensured that the fine-tuning process was both computationally efficient and capable of effectively adapting the LLM to the specific task of PHI annotation.

\section{Synthetic Training Data Generation}
\label{appendix:trainingdata}

During the instruction tuning process, we employed GPT to generate synthetic clinical notes as training data, following two distinct methods: Anonymized Example-Guided Note Generation and Synthetic PHI Insertion into Anonymized Notes, as outlined in Section \ref{sec:syntheticdataapp1}. Specifically, using the Anonymized Example-Guided Note Generation approach, we generated 3,000 synthetic notes, while the Synthetic PHI Insertion into Anonymized Notes approach was used to generate an additional 1,000 synthetic notes.

Initially, we had access to a small set of annotated real-world clinical notes provided by a hospital in the US, from which we selected two representative notes to guide GPT in creating a general structural template. In the Anonymized Example-Guided Note Generation approach, we utilized this structure to generate 3,000 synthetic clinical notes along with their corresponding PHI entities.

For the Synthetic PHI Insertion into Anonymized Notes approach, we extracted clinical information—such as patient gender, allergies, and diagnoses—from the publicly available eICU database to serve as a reference for generating synthetic clinical content. To simulate personal identifiers, GPT generated pools of male and female first names, as well as sets of last names. A first name was randomly selected based on gender, paired with a randomly chosen last name to form a simulated patient identity. Additionally, a 10-digit phone number was randomly generated, and GPT produced random city and street combinations to create realistic addresses. These synthetic personal identifiers were embedded into the clinical notes to mirror real-world documentation. Finally, GPT was tasked with extracting PHI entities from the generated notes to ensure that the synthetic data reflected the diversity and structure of real-world clinical records.

\section{Finetune}
\label{appendix:finetune}
\begin{figure}[h!]
\centering
\includegraphics[width=1\columnwidth]
{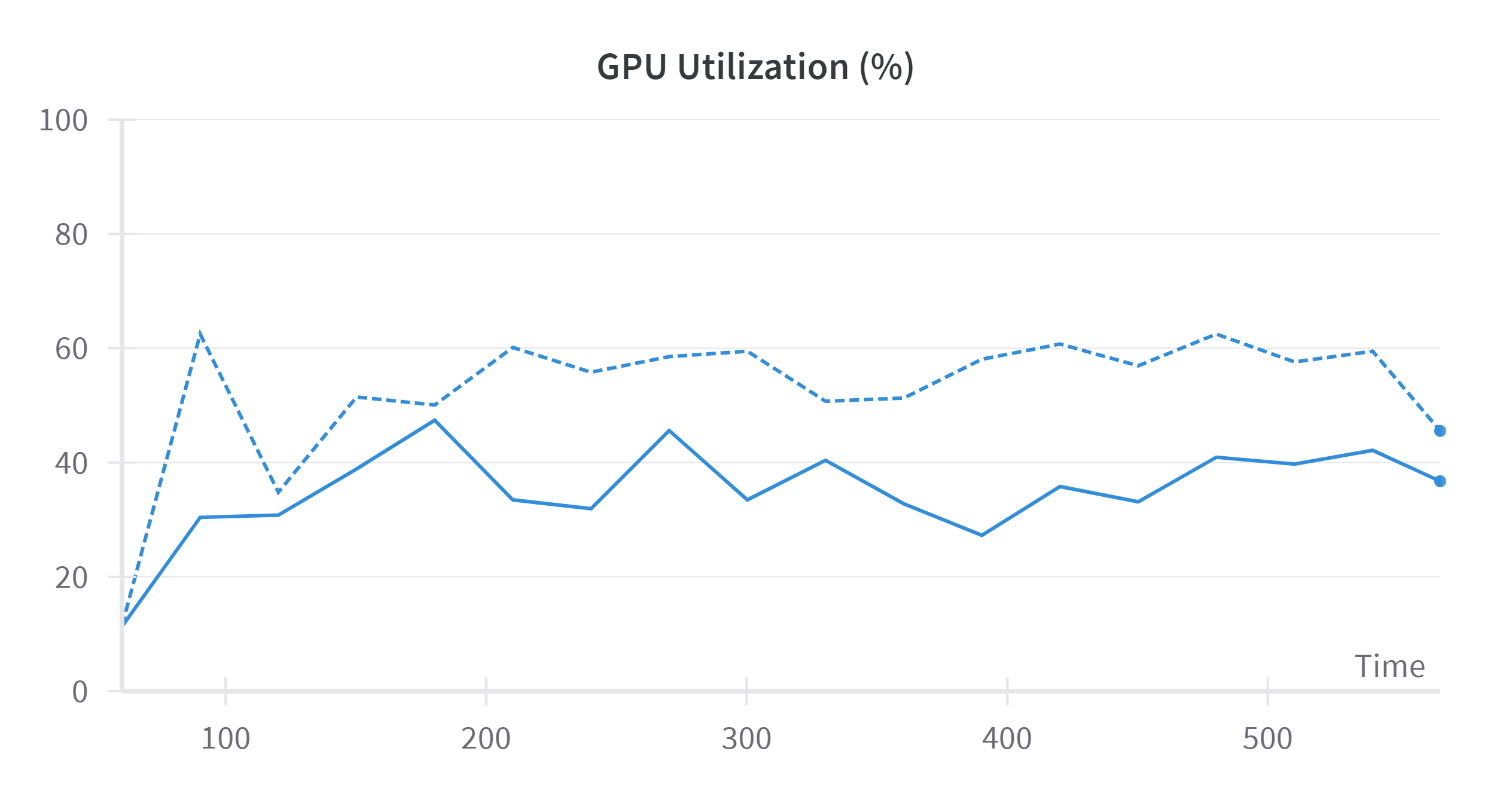}
\caption{GPU Utilization During Fine-Tuning}
\label{fig:5}

\end{figure}

\begin{figure}[h!]
\centering
\includegraphics[width=1\columnwidth]{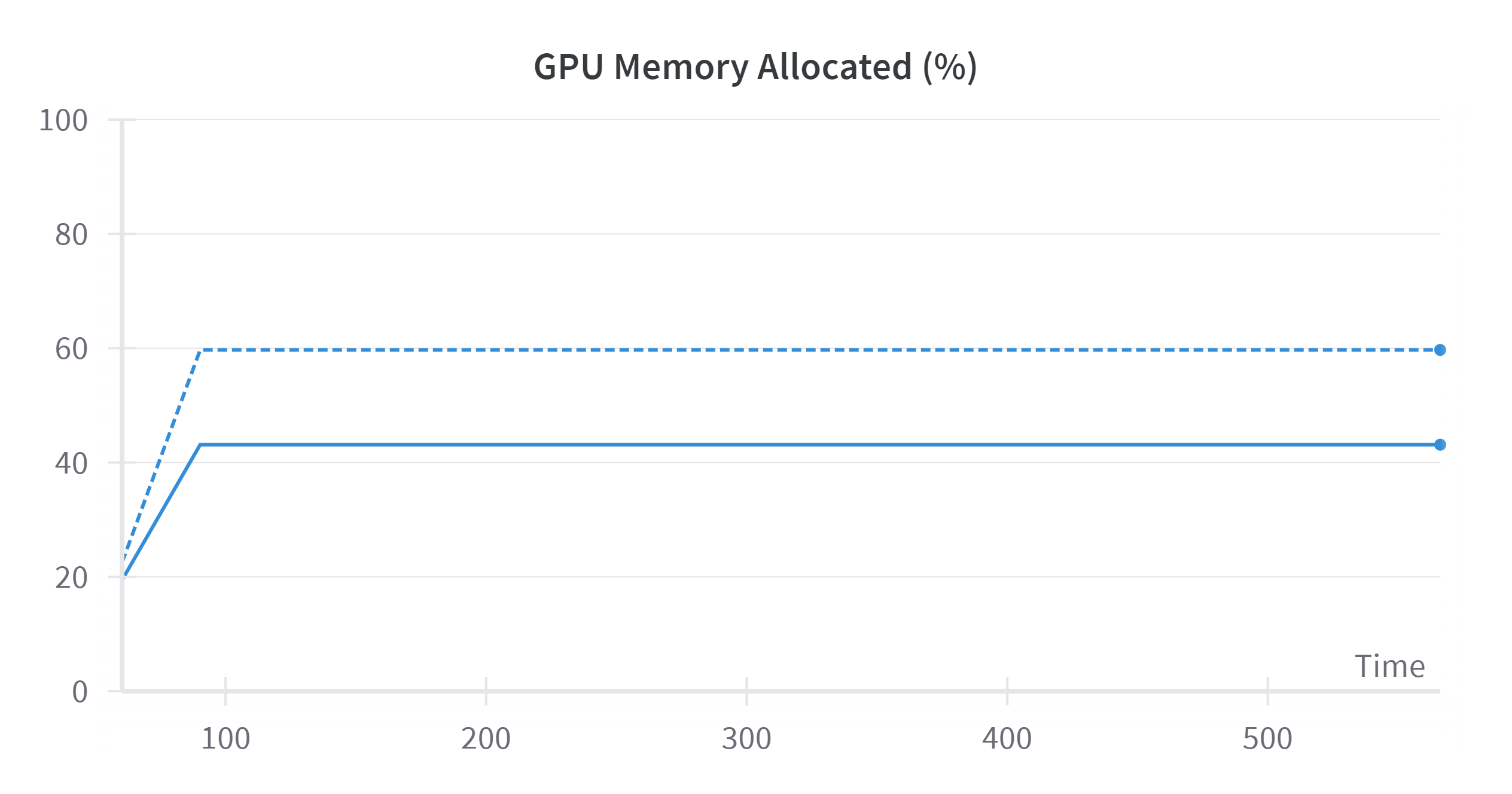}
\caption{Memory Usage During Fine-Tuning}
\label{fig:6}

\end{figure}

In this work, we perform the fine-tuning process on a local server equipped with eight H100 GPUs, and we utilize two of these H100 GPUs. During the instruction tuning process, GPU utilization and memory usage during the fine-tuning process are illustrated in Figure \ref{fig:5} and Figure \ref{fig:6}, respectively.

\section{Synthetic Data Generation Prompt}
\label{appendix:approach2}

\begin{tcolorbox}[colback=blue!5!white,colframe=blue!50!black,title=\small{Anonymized Example-Guided Note Generation Prompt}]
\small 
\textbf{SystemMessage:} ``Act as an experienced doctor. Your goal is to generate simulated clinical notes. A clinical note contains Protected Health Information (PHI), which includes the following entity types: 'PERSON', 'LOCATION', 'ORGANIZATION', 'AGE', 'PHONE\_NUMBER', 'EMAIL', 'DATE\_TIME', 'ZIP', 'PROFESSION', 'USERNAME', 'ID', 'URL'. \\
You are asked to generate simulated clinical notes with PHI information and then extract all PHI entities within the simulated clinical notes and store them in a dictionary. \\
The expected output format is: Clinical Note: Simulated\_Note, PHI: Note\_PHI, where Simulated\_Note is the simulated note, and Note\_PHI is a dictionary containing all PHI elements within the corresponding simulated note. \\
Dictionary Note\_PHI should only include the following keys: 'PERSON', 'LOCATION', 'ORGANIZATION', 'AGE', 'PHONE\_NUMBER', 'EMAIL', 'DATE\_TIME', 'ZIP', 'PROFESSION', 'USERNAME', 'ID', 'URL'. \\
For the 'PERSON' entity type, there are two special cases: 1. When you generate 'Dr. John', you should only extract 'John' as a PHI element; 2. When you generate 'Mr. John', you should take 'Mr. John' as a PHI element.\\
Here are some sample answers I want: \\
Clinical Note: "Chief Complaint: Cardiac Arrest...", PHI: {"PERSON":["John Doe", "Swift"], "ORGANIZATION":["hospital"], "AGE":["24"], "PHONE\_NUMBER":["999-9999-999"]} \\
Clinical Note: "Chief Complaint: Fall...", PHI: {"PERSON":["Jimmy Chen"], "AGE":["30"], "DATE\_TIME":["3/22/2023"]}''\\

\textbf{HumanMessage:} ``Please generate one simulated clinical notes along with a list which contains all Protected Health Information (PHI) entities within the notes.''
\end{tcolorbox}



\begin{tcolorbox}[colback=blue!5!white,colframe=blue!50!black,title=\small{Synthetic PHI Insertion into Anonymized Notes Prompt, Part 1}]
\small 
\textbf{SystemMessage:} ``You are an assistant who helps the doctor write clinical notes.''\\

\textbf{HumanMessage:} ``You are an assistant who helps the doctor write and annotate clinical notes. You should follow the following two steps:\\
1. Please write a clinical note. THE NOTE SHOULD BE AT ABOUT 800 WORDS. Here is some information you can refer to. You MUST use the 'name', 'phone', and 'address' field in PATIENT INFORMATION\\
\textless INFORMATION\textgreater\\
PATIENT INFORMATION:\\
{'gender': 'female', 'age': 69, 'ethnicity': 'caucasian', 'hospitalid': 318, 'wardid': 794, 'admissionheight': 172.5, 'hospitaladmitsource': 'direct admit', 'hospitaldischargestatus': 'alive', 'admissionweight': 63.2, 'dischargeweight': 63.2, 'uniquepid': '021-114154', 'hospitaladmittime': '2101-05-01 11:25:00', 'unitadmittime': '2101-05-01 17:16:00', 'unitdischargetime': '2101-05-07 18:48:00', 'hospitaldischargetime': '2101-05-07 18:48:00', 'name': 'Isla Wilson', 'phone': '958-780-1849', 'address': '5687 Cedar Boulevard, Dallas, TX 75250'}''
\end{tcolorbox}

\begin{tcolorbox}[colback=blue!5!white,colframe=blue!50!black,title=\small{Synthetic PHI Insertion into Anonymized Notes Prompt, Part 2}]
\small 
\textbf{HumanMessage (continue):} ``\\
ALLERGY\\
{'allergyid': 779, 'drugname': 'atenolol', 'allergyname': 'atenolol', 'allergytime': '2101-05-01 17:40:00'}\\
DIAGNOSIS\\
{'diagnosisid': 7965, 'icd9code': '518.81, j96.00', 'diagnosisname': 'acute respiratory failure', 'diagnosistime': '2101-05-04 11:36:00'}\\
LAB\\
{'labid': 145947, 'labname': 'bicarbonate', 'labresult': 32.0, 'labresulttime': '2101-05-01 17:45:00'}\\
MEDICATION\\
{'medicationid': 28775, 'drugname': 'pantoprazole 40 mg inj', 'dosage': '40 mg', 'routeadmin': 'iv push', 'drugstarttime': '2101-05-01 18:00:00', 'drugstoptime': '2101-05-04 16:39:00'}\\
TREATMENT\\
{'treatmentid': 11656, 'treatmentname': 'stress ulcer prophylaxis - famotidine', 'treatmenttime': '2101-05-04 11:36:00'}\\
\textless END OF INFORMATION\textgreater\\

Here is a note as an example:\\
\textless EXAMPLE\textgreater\\
Chief Complaint:  Cardiac Arrest    History Of Present Illness:  $PERSON$ is a $AGE$ y.o. male with no known medical history brought in by EMS following cardiac arrest during intercourse...\\
\textless END OF EXAMPLE\textgreater\\

Note that your output content should be different from the example. Please add the patient's email (The email domain name must be a real one.) and relationship in the clinical note. You can make up the doctor's name, date, patient's email and relationship. You must make up necessary information if they are used.\\

2. After generating the note, extract all PHI entities within the note and store them in a JSON. PHI entity types include: 'PERSON', 'LOCATION', 'ORGANIZATION', 'AGE', 'PHONE\_NUMBER', 'EMAIL', 'DATE\_TIME', 'ZIP', 'PROFESSION', 'USERNAME', 'ID', 'URL'. For 'PERSON' entity type, there are two special cases:\\ 
1. When you generate 'Dr.(Name)', you should only extract '(Name)' as a PHI element; \\
2. When you generate 'Mr./Ms./Mrs.(Name)', you should take 'Mr./Ms./Mrs.(Name)' as a PHI element.\\

Here is an example of PHI:\\ 
\{ "PERSON": ["Emily Turner", "Smith"], 
"AGE": ["28"], 
"ORGANIZATION": ["Midtown Medical Center"], 
"DATE\_TIME": ["September 15th, 2023, at approximately '9:45 PM'"], 
"LOCATION": ["Central Park, New York"], 
"PHONE\_NUMBER": ["555-123-4567"] 
\}\\
(End of Example)\\

Your answer format should be like this:\\
Clinical note: (Your clinical note)\\
PHI: (Your PHI, in JSON format)''
\end{tcolorbox}
\vspace{10mm}




\balance

\bibliographystyle{ACM-Reference-Format}
\bibliography{sample-base}

\end{document}